# Magnetic capsules for NMR imaging: Effect of magnetic nanoparticles spatial distribution and aggregation.

*Azhar Zahoor Abbasi,[a] Lucía Gutiérrez,[b] Loretta L. del Mercato,[a,*] Fernando Herranz,[c] Oksana Chubykalo-Fesenko,[d] Sabino Veintemillas-Verdaguer,[b] Wolfgang J. Parak,[a] M Puerto Morales,[b] Jesús M González,[e] Antonio Hernando[e,f] and Patricia de la Presa.[e,f] †*

a) Philipps Universität Marburg, Fachbereich Physik and WZMW, Renthof 7, 35037 Marburg (Germany)

b) Departamento de Biomateriales y Materiales Bioinspirados, Instituto de Ciencia de Materiales de Madrid/CSIC, Sor Juana Inés de la Cruz 3, Campus de Cantoblanco, Madrid 28049, Spain

c) Instituto de Estudios Biofuncionales, Centro de Investigación Biomédica en Red de Enfermedades Respiratorias (CIBERES). Madrid (Spain)

---

† Corresponding author e-mail: pm.presa@pdi.ucm.es

d) Departamento de Nanoestructuras, Superficies y Recubrimientos, Instituto de Ciencia de Materiales de Madrid/CSIC, Sor Juana Inés de la Cruz 3, Campus de Cantoblanco, Madrid 28049, Spain

e) Instituto de Magnetismo Aplicado (UCM-ADIF-CSIC), P.O. Box 155, Las Rozas, Madrid 28230, Spain

f) Departamento de Física de Materiales, Univ. Complutense de Madrid, Madrid, Spain

**Abstract**

Magnetic and NMR relaxivity properties of $\gamma$-$Fe_2O_3$ nanoparticles embedded into the walls of polyelectrolyte multilayer capsules and freely dispersed in a sodium borate buffer solution have been investigated. The different geometric distribution of both configurations provides the opportunity to study the relationship of water accessibility and magnetic properties of particles on the NMR relaxivity. Changes in their blocking temperature and average dipolar field were modeled as a function of packing fraction in the ensemble of free and entrapped nanoparticles. For free nanoparticles with relatively low concentration, relaxivity values increase with packing fraction according to an increase in the dipolar field and larger water accessibility. However for embedded NPs in the capsule wall, packing fractions should be limited to optimise the efficiency of this system as magnetic resonance imaging (MRI) contrast agent.

**Introduction**

Magnetic nanoparticles (NPs) are interesting contrast agents for Magnetic Resonance Imaging (MRI) thanks to their ability to affect the relaxation rate of water protons, causing a decrease in signal intensity that results in a darkening effect in the corresponding MR image.[1-4] During the past decade an increasing number of works on the use of coated iron oxide NPs as MR contrast have been published (see Ref. 5 and references therein). The effectiveness of MR contrast agents is expressed as relaxivity, which represents the slope of dose-relaxation rate dependencies. As there are two relaxation times ($T_1$ and $T_2$), there are two relaxation rates ($R_1$ and $R_2$) and two relaxivities ($r_1$ and $r_2$). Relaxivities vary with magnetic field strength, temperature and biological environment. Particles synthesized by different procedures exhibit differences in their physicochemical characteristics and hence in their imaging efficacy which is not yet fully established and it will be analyzed in this paper. Magnetic NPs can also be used in cancer therapy, which is known as magnetic field hyperthermia, based on the local heating of the magnetic NPs with an externally applied AC magnetic field and also as drug delivery.[6-10] Another field of interest in biomedicine is the treatment of iron deficiency anemia in humans and farm animals.[11] In addition to their technological significance as diagnostic and therapeutic agents, magnetic NPs are also very interesting for basic research owing to their large surface to volume ratio.[12]

Recently, the scientific interest in the field of magnetic carriers has been shifted from colloidal suspensions of coated iron oxide NPs, forming aggregates sized between 7 and 200 nm, to multicomponent nanocapsules produced by self-assembly of molecular components in which magnetic and/or metal NPs constitute key components.[13-14] General aspects such as particle size, morphology, composition, chemical structure and processing methodology will determine the capsule mechanical properties and finally the performance.[15] New interactions are expected to appear within the different components inside these systems and with the surroundings.[16] For instance, liposomes have been used to encapsulate magnetic NPs, in which particles are confined in the aqueous liposome phase or in the liposomal hydrophobic wall.[17] In both cases by applying an AC magnetic field the whole system could become unstable releasing the therapeutic drug also present in the nanocapsule.[18]

Among the important parameters that influence the relaxivity, the distance between the magnetic particles and external water molecules plays a key role. For example, it has been proven that the presence of a double surfactant will go in detriment of the imaging contrast.[19] Similarly, the proton transverse relaxivity measured for LbL-encapsulated Au-$Fe_2O_3$ and Au-$CoFe_2O_4$ DNA templated nanostructures decreased significantly when compared to the relaxivity for unencapsulated nanostructures.[20] On the other hand, the aggregation of magnetic NPs into micelles has been shown to enhance NMR relaxivity in spite of the fact that water accessibility is reduced due to the reduction in specific surface area.[21-25] Recently, MRI results of FePt NPs encapsulated into the walls of polyelectrolyte multilayer capsules have shown that the relaxivity increases as the FePt concentration increases.[26] It has also been observed that relaxation of water molecules entrapped in a liposome cavity in the presence of magnetic Gd complexes is affected by size and shape of the internal compartment. Thus, liposome osmotic shrinking leads to an enhancement of both relaxation times.[27]

Magnetic properties of NPs, such as the anisotropy and magnetic interactions, have been extensively studied to understand the relaxivity behavior. Another way to characterize these magnetic properties is the measurement of the blocking temperature ($T_B$). $T_B$ defines the temperature at which the majority NPs are blocked at the experiment time window in the minimum state and their magnetization is no sensitive to thermal fluctuations. $T_B$ is related to the most relevant energy barrier in the system that also reflects the particle and the aggregate size. For moderate and large anisotropies, $T_B$ has been observed to decrease as the NP concentration increases. However, for sufficiently small anisotropy values, $T_B$ shifts towards higher temperatures when the particle concentration increases.[28] Experimental data on ferrofluids consisting of magnetite or maghemite NPs show a shift of the $T_B$ towards higher temperatures with increasing concentration. According to that, $r_2$ values increase with aggregate size for maghemite commercial samples such as Endorem, Resovit and Sinerem[29] and for uniform magnetite NPs prepared by thermal decomposition in organic media.[30] It should be remarked that, although relaxivity and $T_B$ depend on same parameters, their relationship is not clear jet and this work tries to enlighten it.

The configuration of iron oxide NPs incorporated into the walls of polyelectrolyte multilayer capsules differs from that of particles embedded into a volume which are virtually equivalent respect to their interaction with water protons. These different configurations offer a unique opportunity for understanding the relationship between water accessibility and intrinsic relaxivity of the entrapped particles which depend finally on their magnetic properties. We compare here magnetic and relaxivity properties of iron oxide NPs embedded in a capsule wall and dispersed in a sodium borate buffer solution (SBBS). Calculation of the average dipolar field and the $T_B$ is performed on the two geometrical layouts in order to understand the different behaviors. The obtained information can help to establish a relationship between the magnetic and relaxometric properties and to design more efficient contrast agents for MRI.

**Experimental section**

*Synthesis of hydrophobic magnetic NPs*

The iron oxide NPs have been synthesized following the procedure reported by Hyeon *et al.*[31] The synthesis yielded nearly monodisperse $\gamma$-$Fe_2O_3$ NPs which were characterized by transmission electron microscopy (TEM) and X-ray diffraction. For more details see the Supporting Information Section I (SI-I). In a next step the hydrophobic NPs were transferred to aqueous solution by coating those with an amphiphilic polymer.

*Synthesis of the amphiphilic polymer*

The synthesis of the amphiphilic polymer and coating of the hydrophobic particles with it were done using a published protocol[32-33] and they are described in SI-II. The resulting hydrophilic particles are dispersed in SBBS and are called FREE_NPs.

*Synthesis of the micro capsules containing hydrophilic γ-Fe$_2$O$_3$ NPs in the wall*

Polyelectrolyte capsules were synthesized using the layer by layer (LbL) technique on calcium carbonate (CaCO$_3$) cores.[34] Those capsules consist of an empty cavity and the magnetic NPs in their polyelectrolyte wall. The wall of the capsules was made using poly(sodium 4-styrenesulfonate) (PSS) as anionic layer, whereas the poly(allylamine hydrochloride) (PAH) and poly(acrylamide-co-diallyl-dimethylammonium chloride) (P(Am-DDA)) were used as cationic layers. In total, two types of polymeric microcapsules were prepared at two different concentrations of γ-Fe$_2$O$_3$ NPs (low and high, named as LCAP and HCAP, respectively). Moreover one sample made of capsules without γ-Fe$_2$O$_3$ NPs was fabricated as control. Schematic illustration of the synthesis of a polyelectrolyte capsule with low and high concentration of γ-Fe$_2$O$_3$ embedded in the wall is shown is Fig. 1. For more details on the encapsulation of γ-Fe$_2$O$_3$ NPs in the wall of microcapsules see SI-III.

*Structural characterization*

The structural characterization was carried out with transmission electron microscopy (TEM), scanning electron microscopy (SEM) and confocal laser scanning microscopy (CLSM). Transmission electron microscopy was performed on JEM 3010 machine operated at 300 kV. Samples for TEM analysis were prepared by putting a drop of low concentrated capsule solution onto a carbon coated TEM grid. The grid was dried in air prior to TEM measurements. SEM measurements were conducted using a JEOL JSM-7500F machine at an operation voltage of 2.0 kV. For SEM analysis, samples were prepared by putting a drop of capsule solution to a glass slide and drying in vacuum. Confocal micrographs were taken with a confocal laser scanning microscope (CLSM 510 META, Zeiss) equipped with a 100x/1.45 oil immersion objective. The capsules were made fluorescent by adding a layer of poly(fluorescein isothiocyanate allylamine hydrochloride) (PAH$_{FITC}$) as second last layer of the wall. The excitation wavelength was 488 nm for PAH$_{FITC}$. The concentration of capsules was directly determined by counting the number of capsules with an optical microscope in phase-contrast mode (see SI-III).

*Fe concentration*

The Fe concentration was then measured with an Inductively Coupled Plasma Optical Emission Spectrometer (ICP-OES) Perkin Elmer Optima 2100 DV. For this purpose samples were digested with nitric acid to oxidize the organic coating and then, with hydrochloric acid to dissolve the particles (see SI-III).

*Magnetic characterization*

The magnetic characterization was performed in a Quantum Design MPMS-5S SQUID magnetometer. The magnetic characterization of the suspensions (0.1 ml) was carried out in special closed sample holders. The characterization consists of hysteresis loops at 5 Tesla and at 5 K and zero-field-cooled and field-cooled (ZFC-FC) curves from 5 to 250 K and 50 Oe applied field. Diamagnetic contribution from water and organic components was evaluated at high fields (larger than 1 T) and subtracted from the experimental data. In the case of ZFC-FC curves, the low value of the applied field (50 Oe) allows discarding the diamagnetic contribution, *i.e.* the magnetic contribution comes mainly from the iron oxide particles. $T_B$ was determined from the maximum of the ZFC curve.

*Magnetic resonance characterization*

Relaxometric properties were also investigated for each aliquot by measuring $T_1$ and $T_2$ protons relaxation times at different dilutions. The relaxation time measurements were carried out in a Minispec MQ60 (Brucker) at 37 ºC and a magnetic field of 1.5 T. From the graph of the Fe-concentration dependent relaxation times, the relaxivities $r_1$ and $r_2$ were determined for each type of sample. A control of hollow capsules was used to correct the relaxivity data.

**Results**

Iron oxide NPs in maghemite phase, $\gamma$-$Fe_2O_3$, were obtained as revealed by X-ray diffraction and confirming previous results (See Fig. SI-3).[31] The mean particle size calculated from TEM images by measuring the size of 240 particles was 11 nm and the standard deviation was 16% (see Fig. SI-2). Particles are spherical and uniform in size, within the monodisperse limit (standard deviation lower than 20%).

Capsules containing different concentrations of $\gamma$-$Fe_2O_3$ NPs were observed under TEM microscopy (Fig. 2). Such analysis confirmed the presence of low (Fig. 2a) and high (Fig. 2b) NPs concentration embedded into the wall of the polyelectrolyte capsules. As expected, capsules collapse after core removal indicating the absence of the $CaCO_3$ cores in their cavities (see also Fig. SI-5). CLSM images of polyelectrolyte capsules with $\gamma$-$Fe_2O_3$ at high and low concentrations are shown in Fig. 3a and 3b, respectively. Fluorescence images of green emitting dye (fluorescein isothiocyanate (FITC)) from the capsule walls can be seen in the first column. The second column shows the spherical shape of hollow polyelectrolyte capsules by optical transmission images. Finally the corresponding overlay of both fluorescence and transmission channels are shown in the third column, showing a perfect match (Fig. 3). The average diameter of capsules obtained from TEM and CLSM images was $3.8 \pm 0.5$ µm. The number of capsules per millilitre was of the order of $10^8$ and it was determined using the microscope in phase contrast mode with a mean capsule diameter of 3.8 µm.

In order to determine the particle concentration per sample and per capsule, iron content was measured by ICP- OES, being 0.93(3), 0.227(7) and 0.053(1) mg/mL for samples FREE_NPS, HCAP and LCAP, respectively. Taking into account that $3 \pm 1 \ 10^4$ Fe atoms are needed to form one $\gamma$-Fe$_2$O$_3$ NP of $\varnothing$ =11 nm and the number of capsules per milliliter in each sample determined with an optical microscope in phase-contrast mode (Figure SI-6) is $3 \ 10^8$, then the number of particles per capsule is $7 \pm 1 \ 10^4$ for LCAP and $3 \pm 1 \ 10^5$ for HCAP. The methodology to determine the number of $\gamma$-Fe$_2$O$_3$ NPs per capsule is described in SI-III.7 and SI-III.8.

The study of the magnetic response and relaxivity of $\gamma$-Fe$_2$O$_3$ NPs in the capsule wall in comparison to those dispersed in SBBS has been done by comparing the samples behavior at the same particle concentration. If the particle concentration is constant, the distance between NPs is much smaller for those in the capsule wall than the ones dispersed in SBBS. To gauge these differences, it is better to define a local concentration or packing fraction.[35] This packing fraction is defined as the total volume occupied by the NPs divided by the volume in which the NPs are distributed. In the case of $\gamma$-Fe$_2$O$_3$ NPs in the capsule wall, the available volume for the NPs is the $4\pi r^2 t$, in which r is the radius of the capsule (r

= 1.9 µm) and t the thickness of the wall. Thickness of one polyelectrolyte bi-layer is around 1.5 nm,[36] the wall of the capsule is made of six bi-layers so that the total thickness of the wall is 18 nm. Therefore, the volume in which the particle can be distributed is 8±2 $10^{-13}$ cm$^3$. The total volume of the NPs was obtained by simply multiplying the volume of one $\gamma$-Fe$_2$O$_3$ NP by the total number of NPs in each sample. Packing fraction for sample LCAP was 6 % and for HCAP was 23 %. On the other hand, FREE_NPs can be distributed in the whole volume of the sample resulting in packing fractions of 0.02 %. As a result, the packing fraction for $\gamma$-Fe$_2$O$_3$ NPs in the capsule wall is 300-400 times larger than that for FREE_NPs at the same Fe total concentration.

The hysteresis loops of $\gamma$-Fe$_2$O$_3$ NPs in the capsule wall and in SBBS do not differ significantly as shown in Fig 4. However, the ZFC-FC measurements show slightly differences for these samples (Fig. 5). The $T_B$ of FREE_NPS decreases for increasing particle concentration, while the contrary occurs for the NPs in capsule wall, in which $T_B$ increases for increasing particle concentration, as shown in Fig 5. Table 1 shows the coercive field and $T_B$ for each sample. Coercivity decreases as the NP concentration decreases independently on the particle spatial distribution.

Relaxivity characterization for particles with different spatial distribution either dispersed in SBBS or in the capsule walls, showed a marked decrease in the relaxivity values ($r_1$ and $r_2$) when particles are encapsulated. The $r_1$ and $r_2$ values go from 3.39 and 43.8 (mM.s)$^{-1}$ for FREE_NPS down to 0 and 1.8 (mM.s)$^{-1}$ for HCAP, respectively. Surprisingly, $r_2$ value increases as the NPs packing fraction in the capsule wall decreases, *i.e.* from sample HCAP to LCAP (Table 1). This trend is opposite to the one observed for FePt NPs in a similar capsule wall, where $r_2$ was observed to increase with NPs packing fraction.[26] However, for free NPs it has been reported that $r_2$ values increase with the aggregate size and therefore with the packing fraction (see Fig. 6).[29,30] The different behavior observed for maghemite NPs encapsulated in the capsule wall may be explained by the different magnetic anisotropy in comparison to FePt and the different distribution geometry when compared to the bibliography data for maghemite/magnetite NPs.

We would like to emphasize that the blocking temperatures of microcapsule is coming from the NPs located at the capsule`s surface, while the relaxivity $r_2$ is defined by the water proton relaxation everywhere in space, not only at the surface of capsules. Therefore, $T_B$ is defined by the magnetostatic field created at the positions of the NPs while $r_2$ is defined by the magnetostatic field created everywhere in space.

**Discussion**

The behaviour of the blocking temperature, which results in low or high values depending on magnetic anisotropy, agglomeration and particle size, can be visualized in a simple model of two magnetic NPs with large angle between their anisotropy axes. When the NPs are far away enough (low packing fraction), they can be considered as two non-interacting magnetic NPs and the $T_B$ is governed by the anisotropy field. As the distance between particles decreases (larger packing fraction), the magnetic interaction increases and produces a drop in the effective anisotropy that leads to a decrease in the blocking temperature. The $T_B$ decreases up to the moment when the magnetic interaction between particles becomes comparable to the anisotropy field. However, when the magnetic interaction is higher than the anisotropy field (even larger packing fraction), NPs behave as a cooperating system and the $T_B$ is ruled by their total volume, *i.e.*, the higher the magnetic interaction, the higher the $T_B$ is. In our case -NPs dispersed in the volume of SBBS or located at the capsule wall- it is necessary to take into account the interactions in a many particle system and also the NPs size distributions; therefore, we perform calculations in this system to calculate the average dipolar field felt by the magnetic particles (affecting the blocking temperature) and the dipolar field created by the magnetic capsules averaged over the overall space (affecting $r_2$ values) as a function of NPs concentration and spatial distribution, *i.e.* packing fraction.

In what follows we present modeling results on ensembles of NPs with different spatial distributions. The complete modeling of the experimental situation is outside of our possibilities since one capsule contains $10^4 - 10^5$ NPs. To get an inside into general tendencies of the magnetic behavior, we consider nanometric capsules, in the range size suitable for in vivo applications. More concretely, we generate an ensemble of NPs with a log-normal distribution having the most probable diameter D=10 nm and 10% of volume dispersion. The NPs are distributed inside a cube of 500x500x500 nm$^3$ in two geometries: (a) uniform volume distribution and (b) uniform distribution within a surface layer of 18 nm on a sphere with D=400 nm diameter (see Fig. SI-7). The NPs are considered non-aggregated. We consider NPs with saturation magnetization values M$_s$= 400 emu/cm$^3$, as corresponding to maghemite. The value of the crystalline anisotropy is not well known. Additionally, in our analysis we consider the macrospin approximation with the effective anisotropy which can be larger than the bulk magnetocrystalline anisotropy due to the surface anisotropy.[37] In the present study we have chosen to consider two possibilities, corresponding to weak and strong interaction cases. In the former case we take the anisotropy parameter corresponding to maghemite NPs as $K$= 4.6 $10^4$ erg/cm$^3$,[37] in the latter case we take the effective anisotropy as $K$= 3.3 $10^5$ erg/cm$^3$, which in the non-interacting case gives the T$_B$ similar to the experimentally measured. The anisotropy axes are considered distributed randomly. For detailed description of the model see the SI-IV.

Our estimation of $T_B$ is based on the evaluation of the energy barrier distribution,[38-39] (see details in SI-IV). The change in energy barrier distributions as a function of the number of NPs per capsule (low packing fraction) is presented in Fig. 7. At low concentrations the main effect is the dispersion of the distribution as a function of the packing fraction. This is easy to understand in terms of the mean dipolar field acting on each NP. The NP with randomly distributed easy axes "feels" a randomly distributed local dipolar field. Depending on its value and the angle with the direction of the easy axis, this field may increase or decrease the NPs largest energy barrier in comparison to the non-interacting NP (Ref. 39). Since the absolute value of the dipolar field is larger for smaller interparticle distances, the dispersion of the energy barrier distribution increases with packing fraction. We define $T_B$ as the temperature at which 90% of NPs are blocked. Figure 8 shows $T_B$ as a function of packing fraction for FREE_NPS sample and capsules for the high anisotropy case. According to the general tendency for the behavior of the energy barrier distributions in Fig. 7, $T_B$ decreases as a function of concentration in both cases, being the $T_B$ values slightly lower for the uniform ensemble than for the capsule systems with the same packing fraction.

In the case considered above the relative strength of the dipolar interactions is not high. Since the energy barrier calculations in a multidimensional space are time consuming we have not been able to take into consideration very high packing densities. To consider the case with strong dipolar fields and high concentrations we present in Fig. 9 the results for a smaller capsule (D=200 nm) and two anisotropy values. At low packing densities the $T_B$ decreases in agreement with the results above. However, at high packing densities the $T_B$ increases. This different regime is due to the fact that at high concentrations the NPs magnetically couple one to another and their energy barriers become collective. Collective energy barriers are associated with volumes of several NPs and thus are large. The transition between the regime of individual energy barriers and collective ones depends on the strength of the dipolar interactions. Thus the corresponding minimum of the $T_B$ versus concentration occurs at larger concentrations in the high anisotropy case. Moreover, the $T_B$ minimum in Fig. 9 is more pronounced for the smallest anisotropy case. This happens due to the fact that the interactions make the energy barrier distributions wider reducing the blocking temperature.

On the other hand, the measured value of $r_2$ is proportional to the strength of the dipolar field acting on the proton of the water in the system. In Fig. 10 we present the absolute value of the dipolar field, averaged over the whole volume, and evaluated for sample FREE_NPS and the capsules, assuming two anisotropy values as described above. The average dipolar field is smaller for the capsules as compared to the uniform distribution, in agreement with the experiment (see the Table 1). This effect has two contributions: (i) In the case of capsules, the average distances between the protons in the water and NPs are larger than for free distributed NPs. (ii) The spherical geometry leads to additional minimization of the dipolar energy, favoring the parallel-to the surface magnetization distribution in systems with high packing density. The latter effect is responsible for the differences observed between high and low anisotropy cases (Fig. 10). In the case of low anisotropy, the magnetostatic interactions are relatively stronger producing additional minimization of the dipolar energy. The difference between the uniform ensemble and the capsule are much more pronounced if the results are re-plotted as a function of the local packing fraction (see Fig. 10b). As an example, for the modeled capsule with 4000 NPs, the packing fraction is around 3% for the uniform ensemble and 23% (as in the experimental sample) on the surface of the capsule. It should be noted that the experimental values of the averaged dipolar field in the case of capsules should be much smaller than the ones obtained in modeling due to a smaller volume concentration of microcapsules.

It should be highlighted the experiment and calculation good agreement in spite of the different capsule size (micro and nanocapsules, respectively). The same tendency for relaxivity and magnetic behavior is observed and allow understanding the role of the NP spatial distribution.

**Conclusions**

To sum up, the absolute value of the dipolar field averaged over the overall space is smaller for the capsule than for the uniformly distributed ensemble leading to a smaller $r_2$ value for the former. The ensembles of FREE_NPs have low packing densities and are in the regime of individual energy barriers; therefore, the $T_B$ decreases as a function of packing fraction. The NPs on capsules have locally large packing densities and their behavior belongs to the regime of collective energy barriers. Consequently, the $T_B$ increases with concentration. The latter is also true for aggregated arrays of NPs as those present in commercial samples, where packing density is high; however, in these cases, the $T_B$ as well as the $r_2$ values increase with concentration, as observed in Fig. 6. Aggregation affects the spatially inhomogeneous particle distribution leading to collective magnetization.

From these results is clear that the geometrical layout of the magnetic NPs plays also a role as important as the magnetic, structural or colloidal properties of the particles (magnetic saturation, anisotropy, homogeneous size and aggregate size). For low NPs concentration relaxivity values are comparable for both geometrical layouts. However, for high NPs concentration, the NPs arrangement on the capsule wall leads to a minimization of the dipolar energy, favoring the parallel-to the surface magnetization distribution. This fact together with lower water accessibility due to the increase in the distance between magnetic capsules at the same Fe concentration, results in lower relaxivity values. The design of complex materials that serve simultaneously as diagnostic and therapeutic agents requires the optimization of each component. Thus, for example, capsules for drug delivery would be better containing large quantities of magnetic NPs to produce an efficient local heating under an AC magnetic field.[40] However, in the light of these results, the NMR imaging contrast produced by this system would be diminished if the particles are encapsulated at high packing density in the capsule wall. NPs with larger magnetic anisotropy could enhance the image contrast.

**Acknowledgements**

AZA is thankful to Higher education commission of Pakistan (HEC) and Deutsche Akademischer Austtausch Dienst (DAAD) for the fellowship. LG and FH hold a Sara Borrel post-doctoral contract. This work was supported by grants from the Spanish Ministry of Science and Innovation (MAT2007-66719-C03-01, MAT2008-01489, MAT2009-14741-C02-00, CSD2007-00010, CS2008-023), the Madrid regional government CM (S009/MAT-1726) and the European Commission (project NAMDIATREAM).

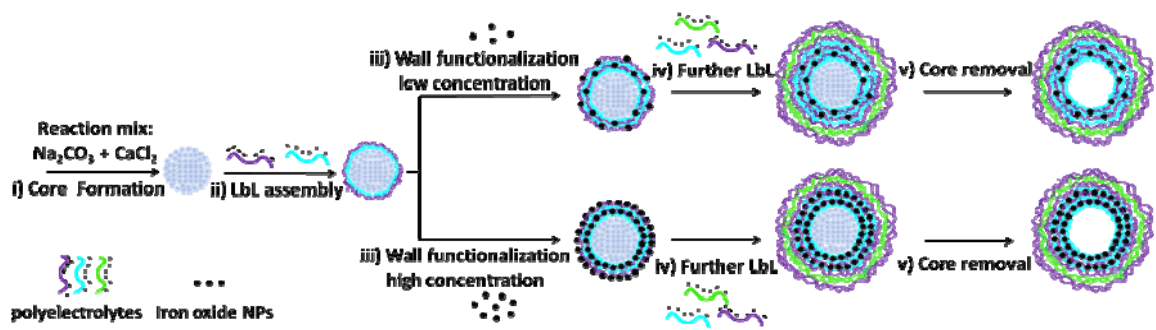

**Figure 1:** Schematic illustration of the synthesis of a polyelectrolyte capsule with low and high concentration of $\gamma$-Fe$_2$O$_3$ embedded in the wall. i) A spherical CaCO$_3$ porous template is synthesized by mixing two solutions of Na$_2$CO$_3$ and CaCl$_2$. ii) The CaCO$_3$ particle is then coated via consecutive LbL deposition of oppositely charged polyelectrolytes to grow a multilayer polymer wall around the template. iii) The wall is functionalized by loading low and high concentrations of charged $\gamma$-Fe$_2$O$_3$ NPs onto an oppositely charged layer during the LbL assembly. The NPs were added as seventh and tenth layer giving the following wall architecture: (PSS/PAH)$_2$(PSS/P(Am-DDA)/$\gamma$-Fe$_2$O$_3$)$_2$(PSS/PAH$_{FITC}$)(PSS/PAH). iv) LbL of polyelectrolytes is repeated to obtain a stable multilayer wall on both capsules with loading low and high concentrations of NPs. A green fluorescent layer was also added to make the wall fluorescent. v) The spherical template is removed to obtain a multilayer capsule with loading low and high concentration of charged $\gamma$-Fe$_2$O$_3$ NPs in the walls. Capsules are not drawn to scale.

| Sample | [Fe] (mg ml$^{-1}$) | H$_c$ (Oe) | T$_B$ (K) | $r_2$ ((mM s)$^{-1}$) |
|---|---|---|---|---|
| **FREE_NPS** | 0.2 | 210 | 45 | 43.8 |
| **FREE_NPS** | 0.05 | 200 | 55 | - |
| **HCAP** | 0.2 | 230 | 65 | 1.8 |
| **LCAP** | 0.05 | 200 | 55 | 46.1 |

**Table 1:** Iron concentration, coercive field (H$_c$), the blocking temperature (T$_B$) and transversal relaxivity values ($r_2$) for maghemite NPs freely dispersed (sample FREE_NP) and entrapped at the capsule wall (HCAP and LCAP). The $r_2$ is determined as a function of concentration at 1.5 T applied field and 37 ºC.

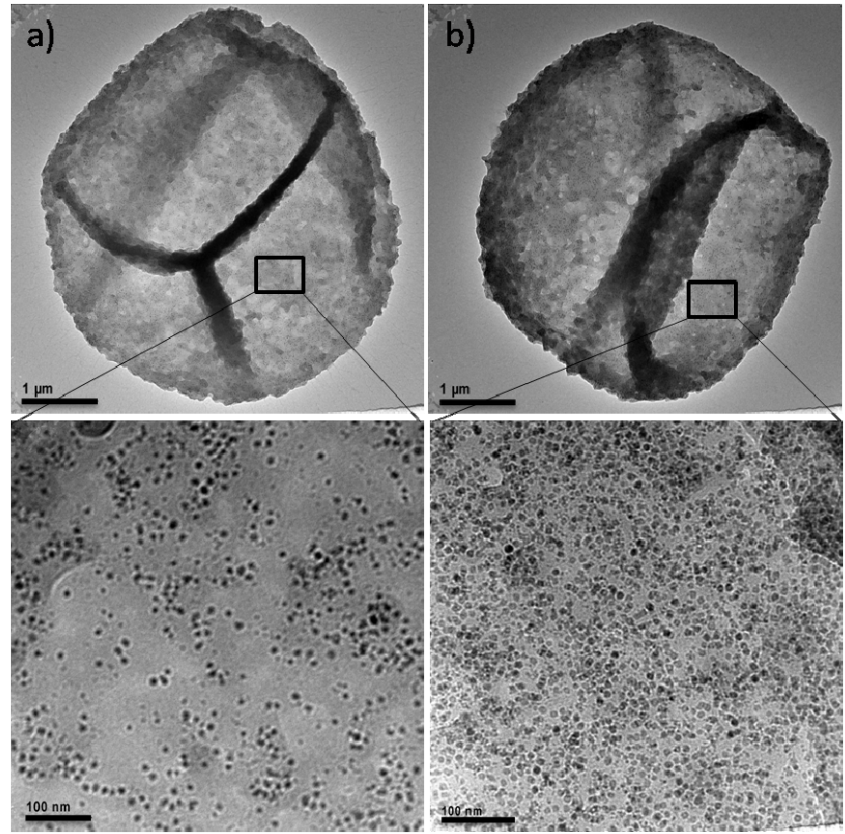

**Figure 2**: TEM images of (PSS/PAH)₂(PSS/P(Am-DDA)/γ-Fe₂O₃)₂(PSS/PAH-FITC)(PSS/PAH) polyelectrolyte capsules after core removal. a) Capsules with low concentration of γ-Fe₂O₃ NPs. b) Capsules with high concentration of γ-Fe₂O₃ NPs. The upper row shows low magnification images of individual capsule. The lower row shows high resolution images zoomed into the capsule wall showing the distribution of the γ-Fe₂O₃ particles in the capsule wall. The scale bars in the upper and lower row correspond to 1μm and 100 nm, respectively.

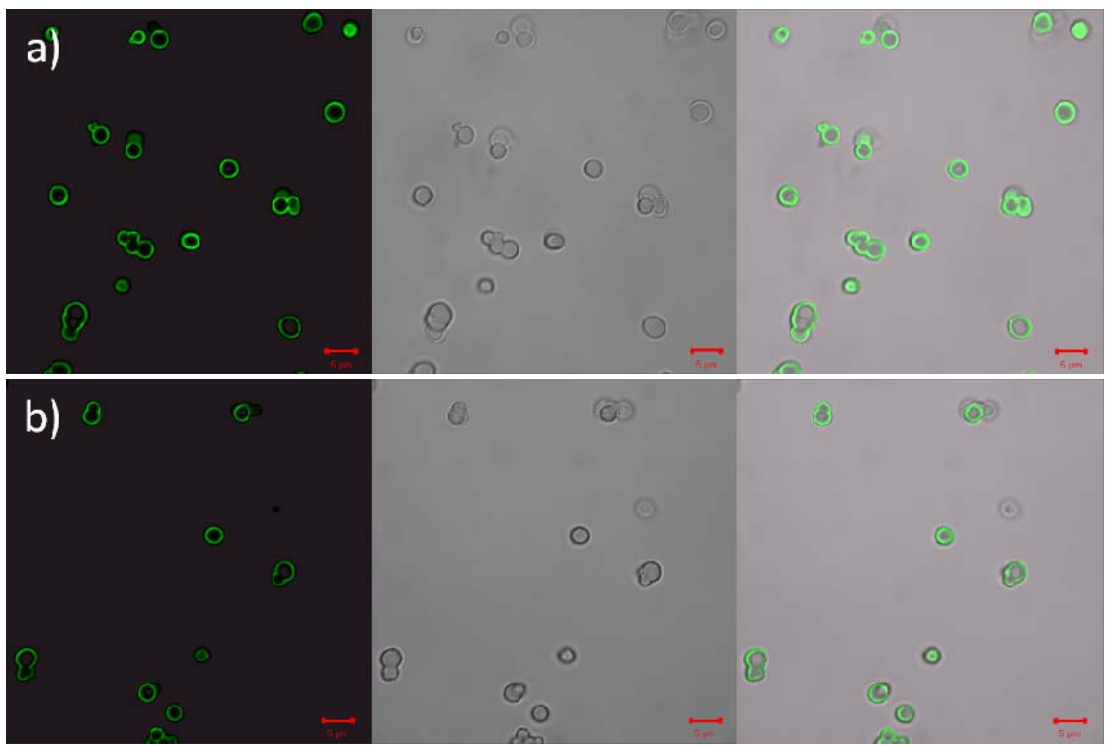

**Figure 3:** CLSM images of (PSS/PAH)$_2$(PSS/P(Am-DDA)/γ-Fe$_2$O$_3$)$_2$(PSS/PAH-$_{FITC}$)(PSS/PAH) poly electrolyte capsules, a) Capsules with low concentration of γ-Fe$_2$O$_3$ NPs. b) capsules with high concentration of γ-Fe$_2$O$_3$ NPs. Left panels: Fluorescence images of green emitting dye. Central panels: optical transmission images. Right panels: corresponding overlay of both fluorescence and transmission channels. Scale bars correspond to 5 µm.

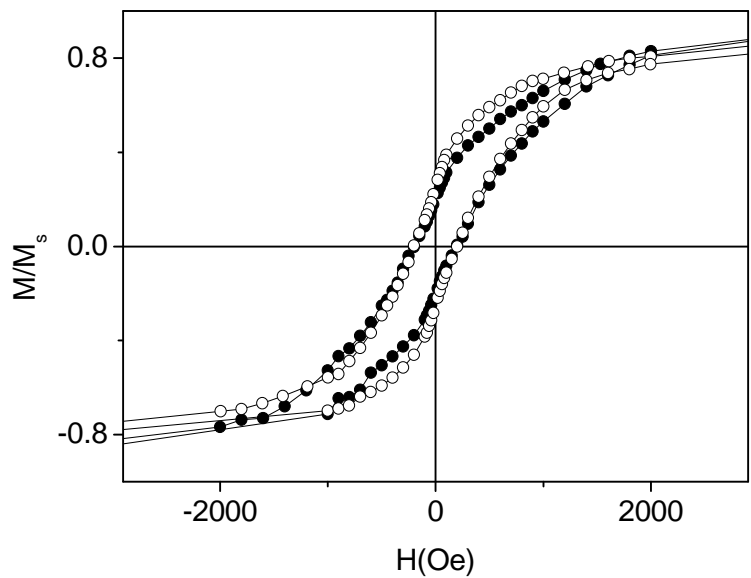

**Figure 4**: Normalized hysteresis curves of γ-Fe₂O₃ NPs in the wall of capsules (full circles) and freely dispersed (open circles).

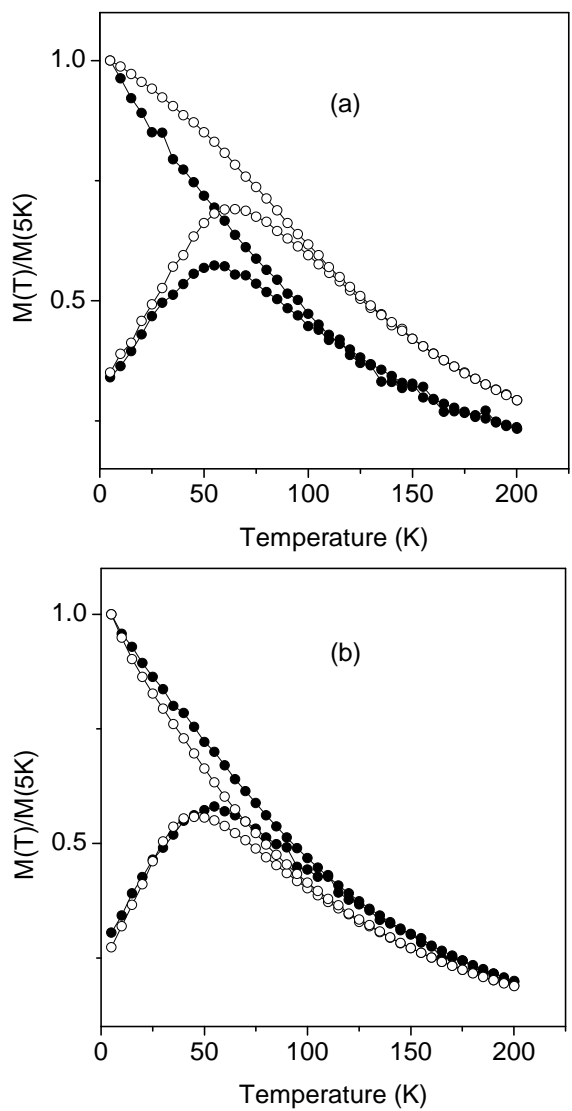

**Figure 5**: ZFC-FC characterization of γ-Fe₂O₃ NPs (a) in the wall of capsules (LbL) and (b) freely dispersed (FREE_NPS) at low (full circles) and high (open circles) concentrations. The curves are normalized to the values at T = 5 K.

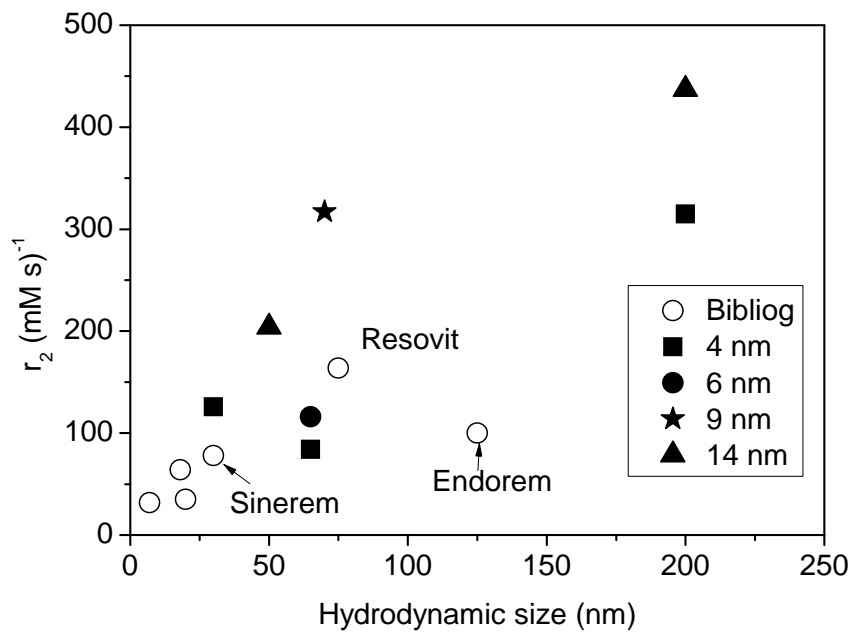

**Figure 6**: Transversal relaxivity values ($r_2$) for magnetic NPs of magnetite with different particle size and aggregate size. (($\circ$) comercial samples Sinerem, Endorem, Resovit from Ref. 26; (■)(●)(▲)(★) data from Ref. 27)

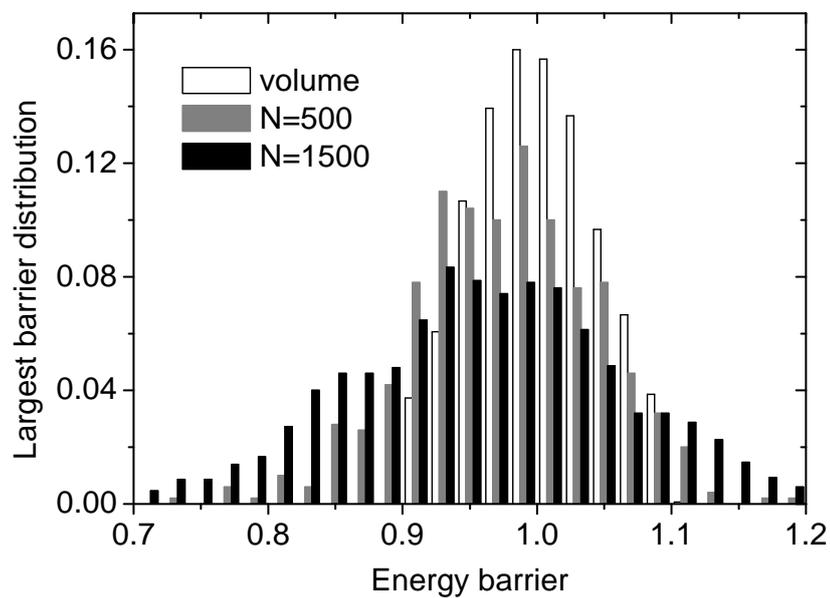

**Figure 7**: Distribution of volumes and energy barriers (normalized to the value K<V>) for a capsule of D=400 nm containing N=500 (packing fraction 3%) and N=1500 (packing fraction 9%) NPs.

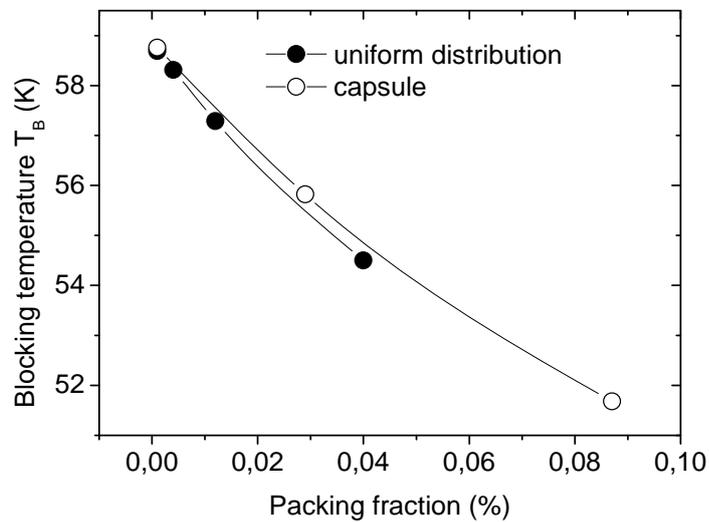

**Figure 8:** Blocking temperatures for an ensemble of uniformly distributed NPs and a capsule with D=400 nm, assuming the anisotropy value K=3.3 $10^5$ erg/cm$^3$ and low values of packing fractions.

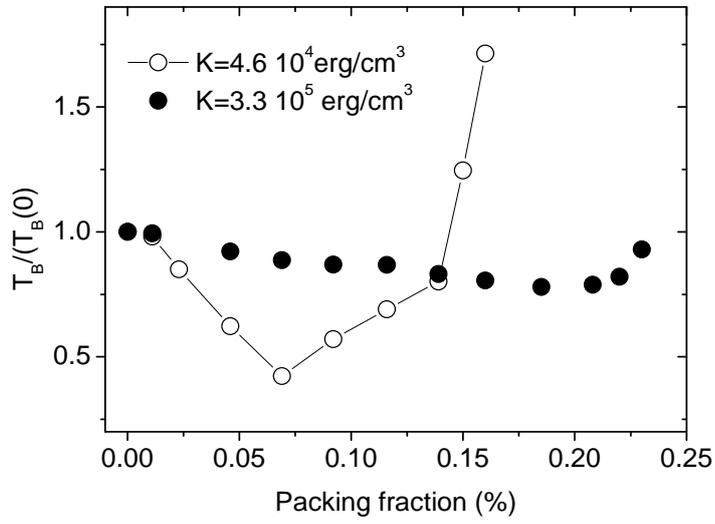

**Figure 9:** Blocking temperature (normalized to the non-interacting ensemble case) as a function of surface packing densities. This calculations are performed for a capsule with D=200 nm and two values of the anisotropy parameter.

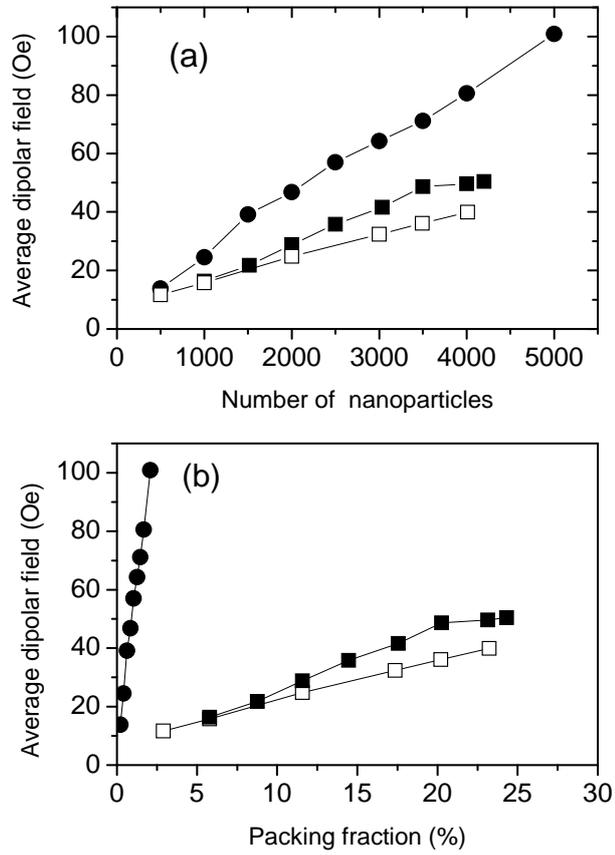

**Figure 10:** Average absolute value of the dipolar field in modeled systems as a function of (a) the number of NPs and (b) the packing fraction. (● free NPs with K= 3.3 $10^5$erg/cm$^3$; ■ capsules with K = 3.3 $10^5$ erg/cm$^3$ and □ capsules with K = 4.6 $10^4$ erg/cm$^3$)

**Supporting Information**

Additional data about the experimental characterization of particles and capsules as well as the numerical model details are available.

**TOC**

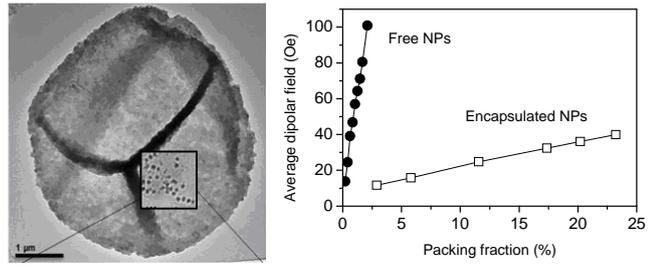